\documentstyle[bo99,epsfig]{article}

\newcommand{\bez}{\begin{eqnarray*}}
\newcommand{\eez}{\end{eqnarray*}}
\newcommand{\be}{\begin{equation}}
\newcommand{\ee}{\end{equation}}
\newcommand{\beq}{\begin{eqnarray}}
\newcommand{\eeq}{\end{eqnarray}}
\newcommand{\bc}{\begin{center}}
\newcommand{\ec}{\end{center}}
\newbox\grsign \setbox\grsign=\hbox{$>$} \newdimen\grdimen \grdimen=\ht\grsign
\newbox\simlessbox \newbox\simgreatbox \newbox\simpropbox
\setbox\simgreatbox=\hbox{\raise.5ex\hbox{$>$}\llap
     {\lower.5ex\hbox{$\sim$}}}\ht1=\grdimen\dp1=0pt
\setbox\simlessbox=\hbox{\raise.5ex\hbox{$<$}\llap
     {\lower.5ex\hbox{$\sim$}}}\ht2=\grdimen\dp2=0pt
\setbox\simpropbox=\hbox{\raise.5ex\hbox{$\propto$}\llap
     {\lower.5ex\hbox{$\sim$}}}\ht2=\grdimen\dp2=0pt

\def\lesssim{\mathrel{\copy\simlessbox}}

\newcommand{\asap}{{A\&A}}
\newcommand{\asaps}{{A\&AS}}
\newcommand{\mnras}{{MNRAS}}
\newcommand{\apj}{{ApJ}}
\newcommand{\apjl}{{ApJ}}
\newcommand{\apjs}{{ApJS}}
\newcommand{\pasj}{{PASJ}}

\newcommand{\spscrev}{{Space Sci. Rev.}}

\def\taut{\tau_{\rm T}}

\def\taumin{\tau_{\min}}
\def\taumax{\tau_{\max}}

\def\sec{\,{\rm s}}

\def\lh{l_{h}}

\def\RXTE{{\it RXTE}}
\def\Ginga{{\it Ginga}}

\def\CGRO{{\it CGRO}}

\title{Time Lags in Compact Objects: \\
Constraints on the Emission Models}

\author{Juri Poutanen}
\affil{Stockholm Observatory, SE-133 36, Saltsj\"obaden, Sweden}

\begin{document}

\maketitle

\begin{abstract}
   Accreting  black holes and neutron  stars in their hard (low) state show
   not only very similar X/$\gamma$-ray spectra but also that the behaviour
   of their light curves is quite similar which can be quantified as having
   similar power-density spectra and Fourier-frequency-dependent time/phase
   lags.  Taken  together  this  argues  for  a  common  mechanism  of  the
   X/$\gamma$-ray  production in these objects.  This mechanism is probably
   a property  of the  accretion  flow only since it does not depend on the
   nature of the compact object.  In this paper, I review the observational
   data paying most attention to the properties of the temporal variability
   such as the time/phase  lags that hopefully can help us to  discriminate
   between  different   theoretical  models.  I  also  discuss  the  models
   developed to account for the basic observational facts.  Particularly, I
   show  that  the  commonly  used  Compton  cloud  models  with   constant
   temperature cannot explain variable sources without violating the energy
   conservation law.  Alternative models where time lags are related to the
   spectral  evolution  during X-ray flares are discussed and compared with
   observations.  Compton  reflection  from the outer edge of the accretion
   disc is shown to markedly affect the time lag Fourier spectrum.
\end{abstract}

\section{Models for the Formation of X/$\gamma$-ray Spectra}
\label{sec:intro}

   X-ray and gamma-ray  spectra of accreting  black holes and neutron stars
   are  deconvolved  into  (at  least)  two  components:  a soft  component
   interpreted as emission from an optically  thick  accretion  disc, and a
   hard tail  associated  with a hot (10--100 keV)  ``corona''.  Reviews of
   the spectral  properties of Galactic black hole candidates (GBHs) can be
   found in Gilfanov et al.  (1995),  Tanaka \& Lewin  (1995),  Grebenev et
   al.  (1993,   1997),   Grove  et  al.  (1997),  and   Poutanen   (1998).
   X/$\gamma$-ray  properties of radio-quiet  active galactic  nuclei (AGN)
   are reviewed by  Zdziarski  et al.  (1997),  Johnson et al.  (1997), and
   Zdziarski (1999).  Recent results on the broad-band spectra of accreting
   neutron stars are presented by Barret et al.  (2000).

   An  amusing  fact is that  super-massive  black  holes in AGN,  GBHs and
   accreting  neutron stars in their hard (low) states (see Tanaka \& Lewin
   1995; Gilfanov et al.  1995 for the  definition of the spectral  states)
   show very similar X/$\gamma$-ray  spectra (see Zdziarski 1999; Barret et
   al.  2000).  Furthermore, properties of their rapid temporal variability
   are also  similar  (van der Klis 1995b;  Wijnands  \& van der Klis 1999;
   Psaltis,  Belloni \& van der Klis  1999;  Ford et al.  1999;  Edelson \&
   Nandra  1999;  Chiang  et  al.  2000).  All  this  argues  for a  common
   mechanism of the X-ray production in all these sources.

   There are good reasons to believe that the main radiative  mechanism for
   the  production  of the hard X-rays is  Comptonization  of soft  photons
   (e.g.,  Shapiro,  Lightman, \& Eardley 1976; Sunyaev \& Tr\"umper  1979;
   Sunyaev \& Titarchuk  1980).  However, it is not  completely  clear what
   determines  the  observed  spectral  slopes.  The  geometry of the X-ray
   emitting  region  and the  source of soft  photons  is still a matter of
   debate (see Svensson 1996; Poutanen 1998; Beloborodov  1999b; Wardzinski
   \& Zdziarski 2000).

   An important clue to our understanding of the X-ray production came from
   the  discovery of Fe lines (at $\sim 6.4$ keV) and the  hardening of the
   spectra  above 10 keV in AGN (Pounds et al.  1990;  Mushotzky,  Done, \&
   Pounds  1993;  Nandra \& Pounds  1994),  Cygnus X-1  (e.g.,  Done et al.
   1992;  Gierli\'nski  et al.  1997), and neutron stars (e.g.,  Yoshida et
   al.  1993).  These features are associated  with the  reflection of hard
   X-rays from cold material (Basko,  Sunyaev, \& Titarchuk 1974; George \&
   Fabian  1991;  Magdziarz  \&  Zdziarski  1995;  Poutanen,  Nagendra,  \&
   Svensson  1996).  These  observations  gave  support  to the  so  called
   two-phase  accretion  disc-corona  models.  In such  models,  X-rays are
   emitted by a hot rarified  corona above the cold accretion  disc (Haardt
   \& Maraschi  1993; Haardt,  Maraschi, \&  Ghisellini  1994; Stern et al.
   1995;  Poutanen \& Svensson  1996).  Hard X-rays from the corona,  being
   reprocessed  in the cold disc,  produce the  reflection  hump as well as
   most of the seed  soft  photons  that are  subsequently  Comptonized  to
   produce  the hard  X-rays.  This is the {\it  feedback}  mechanism.  The
   geometry of the corona  determines the feedback factor which in its turn
   determines   the  spectral   slope  of  the   escaping   radiation.  The
   temperature of the emitting  plasma (or to be more exact, the Kompaneets
   $y$-parameter)  is determined by the energy balance between  heating (by
   magnetic   reconnection?)  and  cooling  (by   Comptonization   of  soft
   photons).

   Further support for the feedback models was recently given by Zdziarski,
   Lubinski, \& Smith (1999) (see also Zdziarski 1999; Gilfanov,  Churazov,
   \&  Revnivtsev  2000) who found a  correlation  between  the  amount  of
   reflection  ($R\equiv\Omega/(2\pi)$,  where  $\Omega$  is a solid  angle
   subtended  by cold  material as viewed  from the X-ray  source)  and the
   intrinsic photon spectral index, $\Gamma$, of the hard X-ray  component.
   Such a correlation  can easily be explained if there is overlap  between
   the hot corona and the cold disc (Poutanen,  Krolik, \& Ryde 1997).  The
   further  the cold disc  penetrates  into the  corona,  the larger is the
   cooling, the smaller is the temperature of the corona, the softer is the
   spectrum,  and, finally, the larger is the amplitude of the  reflection.
   The model, however, appears to have trouble giving reflection amplitudes
   above  $R_{\max}\sim  0.5$ (if the coronal optical depth $\taut \sim 1$,
   see  Zdziarski et al.  1997) due to partial  smearing of the  reflection
   component by the hot corona.

   Alternatively,  the observed $R - \Gamma$  correlation can be reproduced
   by variations of the bulk velocity of the X/$\gamma$-ray emitting plasma
   (Beloborodov 1999a,b).  If the emitting regions are sufficiently compact
   to  produce  electron-positron  pairs,  the  pressure  of the  radiation
   reflected  and  reprocessed  in the disc  accelerates  pairs  to  mildly
   relativistic  velocities  away from the  disc.  On the  other  hand, for
   proton dominated  plasmas, a small anisotropy in the energy  dissipation
   mechanism  can result in the ejection of  particles  away or towards the
   disc.  Ejection away from the disc reduces $R$ below 1 and leads to hard
   spectra, while ejection towards to the disc can result in apparent $R>1$
   as is observed in some objects.

   The physical  possibility of the corona formation was studied by Galeev,
   Rosner, \& Vaiana  (1979).  They showed that the magnetic  fields, being
   amplified  in the cold disc due to  turbulent  motions and  differential
   rotation, do not have time to  annihilate  inside the disc on the inflow
   time  scale.  Instead,  the field  loops are  expelled  from the disc by
   buoyancy (the Parker  instability)  and they  annihilate  in the tenuous
   corona.  Beloborodov (1999a) showed that the mechanism studied by Galeev
   et al.  is able to produce a corona of limited luminosity which is $h/r$
   (the ratio of the disc height to its radius) times smaller than the disc
   luminosity.  By contrast, in some sources most of the energy  escapes in
   the   form  of  hard   X-rays.  Beloborodov   also   argued   that   the
   magneto-rotational   instability  (Velikhov  1959;  Chandrasekhar  1960;
   Balbus  \&  Hawley  1991)  increases  the  rate  of the  magnetic  field
   generation (as compared with the Galeev et al.  model) thus producing an
   active  magnetic  corona  where a large  fraction  of the  gravitational
   energy can finally be dissipated in magnetic flares.  These  qualitative
   arguments  were  recently   supported  by  numerical   three-dimensional
   magnetohydrodynamical  simulations  of Miller \& Stone (2000) who showed
   that  about 25 \% of the  total  energy  dissipation  can  occur  in the
   rarified corona.

   An alternative to the magnetic  corona is the hot disc model (Shapiro et
   al.  1976;  Ichimaru  1977;  Narayan,   Mahadevan,  \&  Quataert   1998;
   Zdziarski  1998; Esin et al.  1998)  which is also able to  explain  the
   observed  X/$\gamma$-ray  spectra.  In order to distinguish  between the
   models, it would be helpful to  compare  the  predictions  of  different
   models with the temporal  variability data (see van der Klis 1995a,b and
   Cui 1999a for recent reviews).  Unfortunately, most of the papers on the
   spectral models do not consider the temporal  variability.  On the other
   hand, most of the models  designed to explain the  temporal  variability
   data do not pay  enough  attention  to the  emission  processes  and the
   physics of the spectral formation.

   In this review, we will  discuss the  variability  data  keeping in mind
   recent  advances  in  modelling  broad-band  X/$\gamma$-ray  spectra  of
   accreting black holes and neutron stars.  Most attention will be paid to
   the time  lags  that  can  shed  light  on the  mechanism  of the  X-ray
   production.  Then we discuss  simple  phenomenological  models  that are
   able to explain some of the observational  facts.  After that, we switch
   to  the  physical   models.  In  particular,   the   properties  of  the
   Comptonizing  regions will be discussed.  We will point out the flaws in
   models that do not consider the energy balance in the ``Compton cloud'',
   and then discuss  models that  satisfy the energy  conservation  law and
   confront them with the available data.

\section{Observing Time Lags in Accreting  Black Holes and Neutron Stars}

\label{sec:lags}

   The standard temporal  characteristics that are usually computed are the
   power-density    spectra   (PDS)   in   different    energy    channels,
   auto/cross-correlation  functions (ACF/CCF), the time/phase lags between
   the variability in different  energy  channels, the coherence  function,
   etc.  Time lags have been studied by two  methods, by  constructing  the
   CCF and by cross-spectral analysis (for details, see Lewin et al.  1988;
   van der Klis 1989; Nowak et al.  1999a).

\subsection{Lags in Black Hole Sources}
\label{sec:lagbh}

\begin{figure}[hbt]
\centerline{\epsfig{file=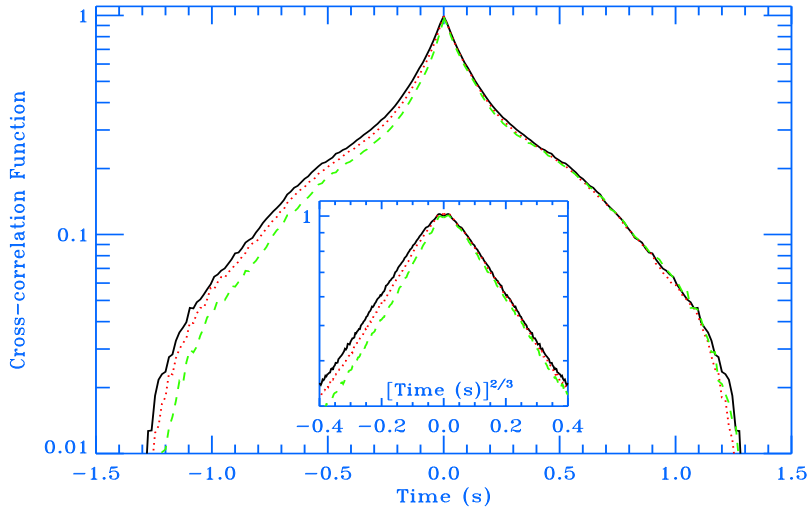,width=11cm,height=7cm}}
\caption{\small 
The cross-correlation functions of Cyg X-1 in the hard state
(\RXTE\ observations from October 22, 1996). The solid curves are the
autocorrelation function for the 2-3.9 keV energy channel,
the dotted and dashed curves are the CCFs for the 6-8.2 and 
14-70 keV vs. the 2-3.9 keV energy channel, respectively.
Note, that CCFs in the hard state sometimes have  much broader wings extending
to $\sim 8\sec$ (e.g., Nolan et al. 1981; Maccarone et al. 2000).  }
\label{fig:ccfcygx1}
\end{figure}

   The  observations  of Cygnus X-1 from sounding  rockets and {\it HEAO 1}
   (Priedhorsky  et al.  1979;  Nolan  et al.  1981)  showed  that  the CCF
   between  different  energy  channels peaks very close to zero lag (delay
   $\lesssim 40$ ms), but it is slightly asymmetric.  Similar asymmetry was
   found in the {\it  EXOSAT}  data by Page (1985) who claimed a $6\pm1$ ms
   shift of the peak of the CCF between the 5-14 keV and the 2-5 keV bands.
   Recent \RXTE\ observations  clearly show asymmetries of the CCFs, which,
   however,    peak    within    $\sim    1$~ms    from   zero   lag   (see
   Fig.~\ref{fig:ccfcygx1}).  This suggests  that the relation  between the
   variation  in the two bands are not simply a time  delay.  Asymmetry  is
   also  observed in the \RXTE\ data of GX 339-4, where the CCFs are offset
   by $\lesssim  5$~ms from zero (using the 2-5 and 10-40 keV bands,  Smith
   \& Liang 1999).  The CCFs of AGN also display similar  properties (e.g.,
   Papadakis \& Lawrence 1995; Lee et al.  1999).

   The CCFs cannot be fitted with simple exponentials at any time scale.  A
   reasonably  good  description  of the  CCF is in  terms  of a  stretched
   exponential $CCF(t)=\beta  \exp[-(|t-t_0|/\tau)^\nu]$  (see the inset of
   Fig.~\ref{fig:ccfcygx1}),  where the normalisation $\beta\approx 1$, the
   time  where the CCFs peak  $t_0<10^{-3}$~s,  $\nu\sim  2/3$ in the range
   $|t|<0.3$~s,  and the time constant  $\tau$ is different  for rising and
   decaying  part of the CCF.  Such a behaviour  is probably  the result of
   self-similarity  of the light curve.  It is interesting to note that the
   ACF of  gamma-ray  bursts  also  show a  similar  stretched  exponential
   behaviour (see Stern \& Svensson 1996; Beloborodov 1999c).

\begin{figure}[hbt]
\centerline{\epsfig{file=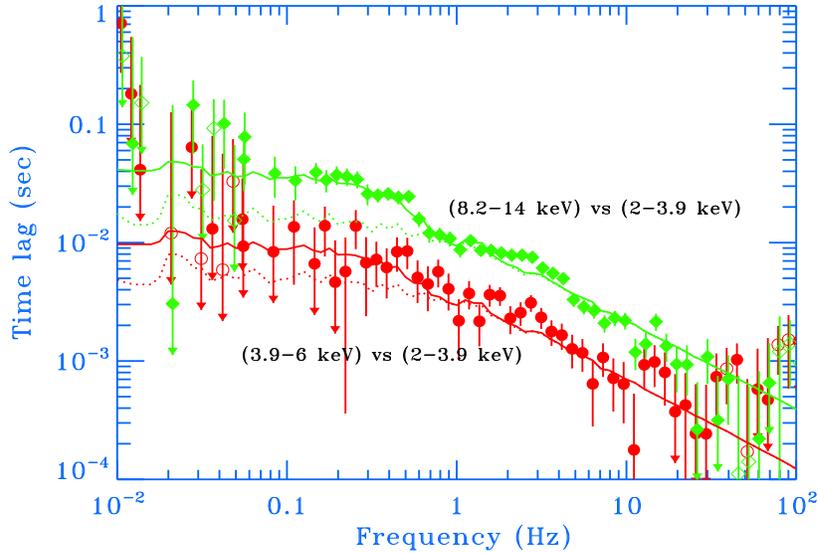,width=11cm,height=7.5cm}}
\caption{\small
   Time lags  between  signals in the 8.2-14 keV and the 3.9-6 keV bands vs
   the 2-3.9 keV band in Cyg X-1 (\RXTE\ observations from October 22, 1996).
   The dotted  curves are the model of Poutanen \& Fabian  (1999b) with the
   maximum flare time scale of  $\tau_{\max}=0.2$~s.  The solid curves show
   the same model with 70\% of the observed Compton reflection  produced at
   a distance of $t_{\rm  refl}\sim 1$ light seconds from the central X-ray
   source.  The  reflector  acts as a low pass  filter  so that  additional
   delays appear only at frequencies  $f\lesssim  1/t_{\rm refl}$, and only
   at  the  energies  where  reflection  is  significant.  See  \S~4.2  for
   details.
}
\label{fig:cygx1lags}
\end{figure}

   Since the CCF does not show  which  frequencies  contribute  most to the
   observed lags, van der Klis et al.  (1987)  suggested to use instead the
   cross-spectrum   for  such  an   analysis.\footnote{The   cross-spectrum
   $C(f)\equiv  S^*(f)  H(f)$,  where  $S(f)$ and  $H(f)$  are the  Fourier
   transforms  of the light  curves in the soft and hard  energy  channels,
   respectively.  The phase lag, $\delta\phi(f)\equiv  \arg[C(f)]$, and the
   time  lag,  $\delta  t(f)\equiv\delta\phi(f)/(2\pi  f)$.  The  lags  are
   positive  when hard  photons are  lagging  the soft  ones.}  In the hard
   state of  Cyg~X-1  observed  by the  \Ginga\  satellite,  the time  lags
   between the  variability  in the 1.2-4.7 and  4.7-9.3  keV energy  bands
   reached  0.1 s and had a  strong  Fourier-frequency  dependence  $\delta
   t(f)\sim  f^{-1}$,  i.e., the  phase  lag  $\delta\phi(f)\approx$  const
   (Miyamoto et al.  1988;  Miyamoto \& Kitamoto  1989).  Similar lags were
   observed in other GBHs, GX~339-4 and GS~2023+338  ($\equiv$V404 Cyg), in
   their hard state  (Miyamoto  et al.  1992).  The  analysis of the \RXTE\
   data for Cyg X-1 (Nowak et al.  1999a),  GX~339-4 (Nowak et al.  1999b),
   1E~1740.7-2942  and  GRS~1758-258  (Smith et al.  1997), and GS 1354-644
   (Revnivtsev et al.  2000)  confirming  the general  features seen in the
   \Ginga\ data, showed more complicated behaviour of the phase lag spectra
   which    have    a    number    of    shelves     and    breaks     (see
   Fig.~\ref{fig:cygx1lags}).

   Grove et al.  (1998) extended this analysis to higher  energies with the
   data from  \CGRO/OSSE.  The time lags at low  frequencies  reached 0.3~s
   between  the 50-70 and the  70-100  keV  photons  in the light  curve of
   GRO~J0422+32  ($\equiv$Nova  Persei  1992).  The breaks  detected in the
   time  lag  spectrum  at 0.1 Hz may  be  related  to  the  quasi-periodic
   oscillation  (QPO) observed at 0.23 Hz.  The \CGRO/BATSE data of Cyg~X-1
   (Crary et al.  1998),  GRO~J0422+32 and GRO~J1719-24  ($\equiv$Nova  Oph
   1993)  (van der  Hooft  et al.  1999a,b)  show  very  similar  time  lag
   spectra.

   The time lags of GBHs in their soft  state  (when  $\Gamma\approx  2.5$)
   turn  out to be  somewhat  different.  {\it  Soft}  lags  were  observed
   between the 1.2-2.3 keV and the 2.3-4.6 keV bands in GX~339-4  (Miyamoto
   et al.  1991) and GS~1124-68  ($\equiv$Nova Muscae 1991; see Miyamoto et
   al.  1993; Takizawa et al.  1997), while the higher energy  photons were
   lagging the variability in the 2.3-4.6 keV band.  In one  observation of
   GS~1124-68,  the  variability  in the 4.6-9.2 keV was the most advanced.
   The phase lag reached $\sim 1$ rad which is much larger than the lags in
   the broad band noise  observed  in the GBHs in their hard  state.  Rapid
   time variations were mostly due to the harder power-law  component which
   is clearly seen from the rms amplitude.  It is interesting that the lags
   seem to  saturate  above 10 keV.  It is  worth  pointing  out  that  the
   largest lags here are observed at the QPO frequency.  On the other hand,
   the time lag  spectrum of Cyg~X-1 in the soft state looks quite  similar
   to that in the hard state (Cui et al.  1997).

   Time lags have been  observed in  GRS~1915+105  in the 67 mHz QPO by Cui
   (1999b)  and in the  broad-band  noise  and QPOs by Reig et al.  (2000).
   The lags show a very  complicated  structure,  sometimes  changing signs
   from one harmonic to another, and the sign also depends on the frequency
   of the QPO.  Wijnands,  Homan, \& van der Klis (1999) and Cui, Zhang, \&
   Chen  (2000)  observed  lags in the  broad-band  noise  and the  QPOs of
   XTE~J1550-564.


\subsection{Lags in Neutron Star Sources}

    Hard time lags in neutron  star  sources  were  discovered  by Hasinger
   (1987) in the CCF of Cyg X-2  (comparing the 1-5 and the 5-17 keV bands)
   in its horizontal  branch (HB, see Lewin et al.  1988 for definitions of
   branches).  The CCF also had  sinusoidal  oscillations  due to the  QPO.
   The time lags showed  anti-correlation  with the QPO frequency, dropping
   from 4 ms to 1.5 ms when the QPO changed  from 20 Hz to 50 Hz.  Hasinger
   interpreted the lags as delays due to scattering (Comptonization) in the
   hot cloud and the  anti-correlation  as an indication of a change in the
   system size.  Associating the QPO frequency with the Keplerian frequency
   at  some   radius   gives   the   relation,   $\delta   t  (f)   \propto
   f_{QPO}^{-2/3}$,  while the actual  data are much  better  described  by
   $\delta t \propto f_{QPO}^{-1}$, i.e.  $\delta\phi = 2\pi f \cdot \delta
   t={\rm const}$.

   Using  the  cross-spectrum   techniques  van  der  Klis  et  al.  (1987)
   confirmed  the existence of $\sim 3$ ms hard lags in the 20-40 Hz QPO of
   Cyg X-2  (and GX  5-1)  and  discovered  8 ms  {\it  soft}  lags  in the
   low-frequency  noise,  which  were  interpreted  as a  softening  of the
   spectrum during the shots that cause QPOs.  These results were confirmed
   by Vaughan et al.  (1994)  who also  showed  (from the  analysis  of the
   \Ginga\  data of GX 5-1 on the HB) that  the  time  lags  increase  with
   photon energy.

   In both Cyg X-2 and GX 5-1 in their normal branch, the lags at the $\sim
   5$ Hz QPO showed energy  dependence  (Mitsuda \& Dotani 1989; Vaughan et
   al.  1999)  reaching  $\delta\phi  \sim \pi$ rad  (i.e.,  $\delta  t\sim
   0.2\sec$)  for 10 keV  photons  vs 2 keV.  At the  same  time,  the  rms
   amplitude  of the QPO in Cyg X-2 had a minimum at 5 keV and in GX 5-1 it
   increased  above  2.5  keV.  This  behaviour  can  be  interpreted  as a
   pivoting of the spectrum around 3-5 keV.

   With the larger  effective  area of \RXTE, Ford et al.  (1999) and Olive
   \& Barret (2000)  discovered phase lags in the broad-band noise of three
   atoll sources, 4U0614+09, 4U1705-44, and 4U1728-34.  These lags are very
   similar to those in GBHs like Cyg~X-1 and GX~339-4,  which tells us that
   the mechanism  responsible  for the lags does not depend on the presence
   or absence  of the hard  surface  of the  neutron  star,  magnetosphere,
   boundary layer, etc., but instead is a property of the accretion flow.

   A number  of  neutron  stars  show kHz QPOs in  their  light  curves  as
   revealed  by \RXTE.  Kaaret  et al.  (1999)  find 25  $\mu$s  soft  lags
   between  the 4-6 keV and the $>$ 9 keV  photons in the 800 Hz QPO in the
   atoll source  4U1636-536.  Analysing the 550 Hz  oscillations  of Aquila
   X-1,  Ford et al.  (1999)  found  soft  lags,  $\delta\phi\sim  1$  rad,
   between the 3-6 keV and the $>$ 6 keV photons.  Similar  lags were found
   in the accreting  millisecond  pulsar SAX J1808.4-3658  (Cui, Morgan, \&
   Titarchuk  1998;  Ford  2000).  The lags in the 830 Hz QPO in 4U 1608-52
   (Vaughan et al.  1997, 1998)  reach 60 $\mu$s  between  the 5 and the 25
   keV photons.

\section{Phenomenological Energy Dependent Shot Noise Models}

\label{sec:shot}

   It is clear that the observed  zoo of time lags cannot be  explained  by
   any single model.  Different  mechanisms should be involved in producing
   the lags at different Fourier  frequencies, the lags in the QPOs and the
   coherent pulsations, and the lags in the broad-band noise.

   Let us first  consider the simplest  possible  model that  produces time
   lags:  a shot noise model  (Terrell  1972),  where shots  (=flares)  are
   uncorrelated  with each other.  We assume that the shot time profiles at
   different  energies  have the same shape, but  slightly  different  time
   constants.  As an  example  we  take a shot  profile  at  soft  energies
   $s(t)=[t/\tau]^p     \exp[-t/\tau]$     and     at     hard     energies
   $h(t)=[t/(\eta\tau)]^p  \exp[-t/(\eta\tau)]$,  where  $t>0$ is  measured
   from  the  beginning  of the  shot  and  $p$ is  positive.  The  Fourier
   transforms,     $S(f)$    and    $H(f)$,    are    $S(f)\propto     \tau
   \Gamma(p+1)/(1-i2\pi    f   \tau)^{p+1}$   and   $H(f)\propto   \eta\tau
   \Gamma(p+1)/(1-i2\pi   f   \eta\tau)^{p+1}$.  The  PDSs   are   $\propto
   |S(f)|^2$ and $|H(f)|^2$.  For small frequencies, $f\ll 1/(2 \pi \tau)$,
   the PDSs have a flat  dependence on frequency,  $\propto f^0$, while for
   large   frequencies,   $f\gg  1/(2  \pi   \tau)$,   the  PDSs  decay  as
   $f^{-2(p+1)}$.  The power per  logarithm of  frequency  (i.e.,  $f\times
   PDS(f)$)   peaks   for   soft   photons   at   $f_{s,\max}=1/[   2   \pi
   \tau\sqrt{2p+1}]$  and at  $f_{h,\max}=1/[\eta  2 \pi \tau \sqrt{2p+1}]$
   for hard  photons.  The phase  lags  $\delta\phi(f)=(p+1)  [\arctan(\eta
   2\pi  f\tau)-\arctan(2\pi  f\tau)]$.  The  lag  is  positive  when  hard
   photons  are  lagging  soft  ones  (i.e.,  for  $\eta > 1$).  For  small
   frequencies,  $\delta\phi(f)$  rises as $\approx  (p+1)  (\eta - 1) 2\pi
   f\tau$, while for large frequencies, it decays as $(p+1)(\eta  -1)/(\eta
   2\pi      f\tau)$.     The     lag      reaches     a     maximum     of
   $\delta\phi_{\max}=2(p+1)(\arctan  \sqrt{\eta}  -\pi /4)$ at  $f=1/(2\pi
   \tau   \sqrt{\eta})$   close  to  $f_{s,\max}$  and   $f_{h,\max}$,  the
   frequencies where $f\times PDS_{s,h}(f)$ peak.

   One can also  consider a modified  shot noise model, where the shot time
   scales are  distributed  according  to a power  law,  $\rho(\tau)\propto
   \tau^{-p}$  between  $\taumin$  and  $\taumax$  (see, e.g.,  Miyamoto \&
   Kitamoto 1989; Lochner, Swank, \& Szymkowiak  1991), with the same ratio
   $\eta$.  Physically this could correspond to, for example, the situation
   when flares of different  durations  appear at different  radii from the
   central object (Poutanen \& Fabian 1999a).  A power-law  distribution of
   $\tau$  assures  that the PDS is also a power-law  $\propto  f^{-(3-p)}$
   (Lochner et al.  1991).  If the flares are self-similar,  then the phase
   lag   will   be   constant   $\approx   \delta\phi_{\max}$   for   $f\gg
   f_{\min}\equiv  1/(2\pi  \taumax)$  and  $f\ll  f_{\max}\equiv   1/(2\pi
   \taumin)$,  decay  as  $1/f$  at  $f>f_{\max}$   and  rise  linearly  at
   $f<f_{\min}$.  The  corresponding   time  lags  are  constant,   $\delta
   t_{\max}=2\pi  \taumax  \delta\phi_{\max}$,  for $f<f_{\min}$, and decay
   approximately as $1/f$ between  $f_{\min}$ and $f_{\max}$.  Note that in
   this model the coherence  function  (Vaughan \& Nowak 1997; Nowak et al.
   1999a) is close to unity, since the light curves at  different  energies
   are almost perfectly synchronised.

   If we  assume  that  $\eta>1$,  there  are hard  lags and the  predicted
   behaviour  of the time lags and  coherence  function  is in a very  good
   agreement  with the  observations  of GBHs  (Poutanen \& Fabian  1999a).
   However,   this   model   contradicts   the   CCF  of   Cyg   X-1   (see
   Fig.~\ref{fig:ccfcygx1}  and  Maccarone  et al.  2000).  The CCF becomes
   narrower at larger  energies which  requires the shots to be narrower at
   larger  energies.  If one,  however,  reverses the time  profiles of the
   shots,   so  that   they   rise   slower   and   decay   faster   (e.g.,
   $s(t)=(-t/\tau)^p\exp(t/\tau),  t<0$), and one assumes  $\eta<1$  (i.e.,
   hard shots are narrower), the CCFs and the time lags can be  reproduced,
   simultaneously.

   Much more  complicated  models  which also  account for lags in the QPOs
   sources were  developed by Shibazaki  et al.  (1988).  We just note here
   that if a signal consists of shots appearing almost  periodically and if
   shots at different energies are shifted in time one against another (or,
   e.g., the minima  are  reached  at the same time and the peaks are not),
   the phase lag has a very  complicated  dependency  on  frequency  (e.g.,
   changes  sign  from one  harmonic  to  another)  depending  on the  shot
   profiles.

\newpage

\section{Physical Mechanisms for Producing Lags}

\subsection{Static Compton Cloud Models}
\label{sect:comp}

   Since Comptonization is the most probable mechanism for X-ray production
   in compact  objects, it is natural to attribute the time delays  between
   hard and soft photons to this  process.  Hard  photons are the result of
   more  scattering  and so  emerge  after,  or lag  behind,  softer  ones.
   Consider a static  ``Compton  cloud'' with fixed Thomson  optical depth,
   $\taut$,  and  electron  temperature  $\Theta=kT_e/m_ec^2$.  A soft seed
   photon of energy $E_0$ injected into the cloud increases its energy by a
   factor of  $A_1=1+4\Theta+16\Theta^2$  on average after each scattering,
   so that after  $N$-scatterings  its energy  $E_N=A_1^N E_0$.  The photon
   mean free path is  $\lambda\approx  R/\max(1,\taut)$  (where  $R$ is the
   size of the X-ray producing  region, and where we accounted for the fact
   that  we are  interested  only  in  those  photons  that  actually  have
   undergone  scatterings  in  the  cloud).  The  time  between  successive
   scatterings  is then  $t_c=R/(c  \max[1,\taut])$,  so the time needed to
   reach the energy  $E_N$ is  (Sunyaev  \&  Titarchuk  1980;  Payne  1980)
   $$t_N=Nt_c=\frac{R/c}{\max(1,\taut)}  \frac{\ln(E_N/E_0)}{\ln  A_1}  ,$$
   which  translates  to  $t_N\sim  10^{-4}\sec$  for  $kT_e\sim  50$  keV,
   $\taut\sim 1$, $R=10$ km, and $E_N/E_0\sim 10$.

   This model was  criticised  by Miyamoto et al.  (1988),  Miyamoto et al.
   (1991) and  Vaughan et al.  (1994).  First, the large  size of the cloud
   ($10^3-10^5 R_g$, where $R_g$ is the Schwarzschild radius, $2GM/c^2$) is
   needed to produce large delays observed in GBHs and neutron stars.  Such
   cloud is physically  unrealistic, since most of the gravitational energy
   is dissipated  within  $10R_g$.  Second, the lags predicted by the model
   are  independent  of  the  Fourier  frequency  (Miyamoto  et  al.  1988)
   contrary to the  observed  $\sim 1/f$  dependence.  Third, it is assumed
   that the soft photons  produce the  variability,  while the hot cloud is
   not variable.  Observationally,  it is well established that when a soft
   black body  spectrum is observed in GBHs it is much {\it less}  variable
   than the hard  X-rays  (e.g.  Miyamoto  et al.  1991),  so that the hard
   X-ray variability is most probably intrinsic to the hot cloud itself.

   Finally,  we would  like to point  out that  due to the  requirement  of
   energy  conservation  the whole concept of a static Compton cloud with a
   constant temperature is physically unrealistic.  The total emitted X-ray
   luminosity  (produced by  Comptonization  of soft  radiation) is $L_{\rm
   tot}=L_h+L_s$,  where  $L_h$ is the  heating  rate in the hot  cloud and
   $L_s$ is the  luminosity  of seed soft  photons.  For hard spectra (i.e.
   $L_h\gg L_s$), $L_{\rm  tot}\approx L_h$.  The total X-ray luminosity is
   thus a function of the heating rate {\it only} and it does not depend on
   the amount of seed soft  photons.  By  changing  $L_s$, one  effectively
   changes the spectral slope of the emergent  X-ray  radiation  which is a
   function of the Compton  amplification  factor $A\equiv  L_h/L_s$.  This
   results in the pivoting of the spectrum (see Poutanen 1998;  Beloborodov
   1999b) without a noticeable  increase in $L_{\rm  tot}$.  The larger the
   $L_s$,  the  smaller  the   equilibrium   temperature  of  the  emitting
   electrons, and the softer the spectrum.  By contrast, in static  Compton
   cloud models no changes in the electron  temperature  are  considered in
   reaction to the changes in the number of soft  photons,  violating  thus
   the  energy   conservation   law.  In  order  to  increase  the  emitted
   luminosity, one has to change the energy  dissipation rate in the cloud,
   but then exactly these changes will be driving the variability.

   Recently,  Kazanas,  Hua, \& Titarchuk  (1997) (see also  B\"ottcher  \&
   Liang  1998; Hua,  Kazanas,  \& Cui 1999)  modified  the  Comptonization
   model.  Instead of a  homogeneous  Compton  cloud, the density  profile,
   $n(r)\sim  1/r$, was assumed (in this case, one has equal  optical depth
   per logarithm of radius).  The  variability  is still driven by changing
   the rate of soft  photon  injection  in the  center of the  cloud.  Then
   larger radii produce lower  frequency  variability  (filtering  out high
   frequency  signal) and larger  lags,  while the  smaller  radii  produce
   higher  frequency  variability  and smaller  lags (see also Nowak et al.
   1999a,c).  This model  solves only one of the  problem  mentioned  above
   ($1/f$ time lag dependence),  while the other problems remain  unsolved.
   Another  modification of the model was considered by B\"ottcher \& Liang
   (1999) based on an earlier suggestion by Miyamoto et al.  (1988).  Here,
   small cold clouds are assumed to  free-fall  into the hot central  cloud
   thus changing the input of soft photons.  Again, this model has problems
   with energy conservation.

\subsection{The Dynamic Compton Cloud}

\begin{figure}[hbt]
\centerline{\epsfig{file=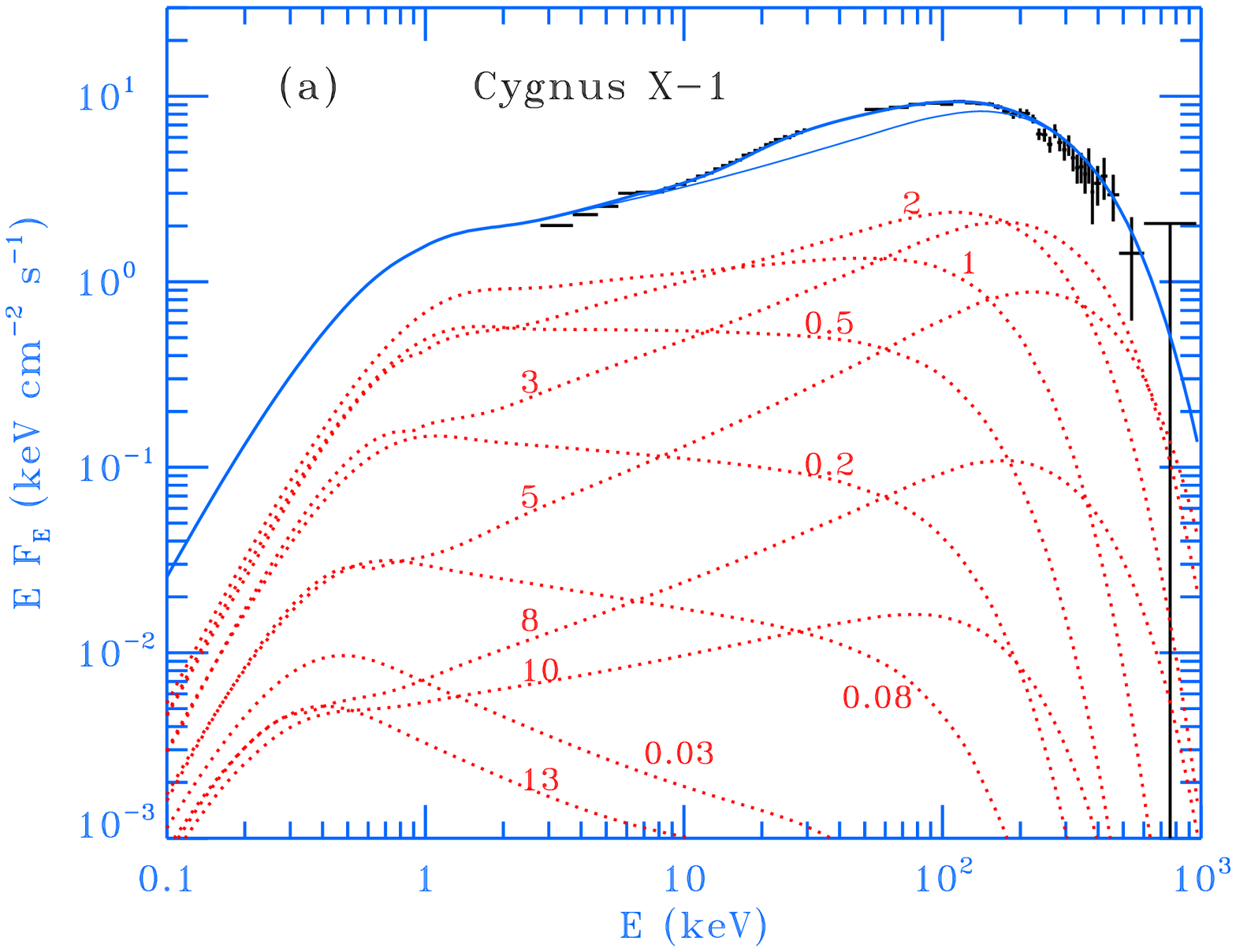,width=6.3cm,height=6cm}
\epsfig{file=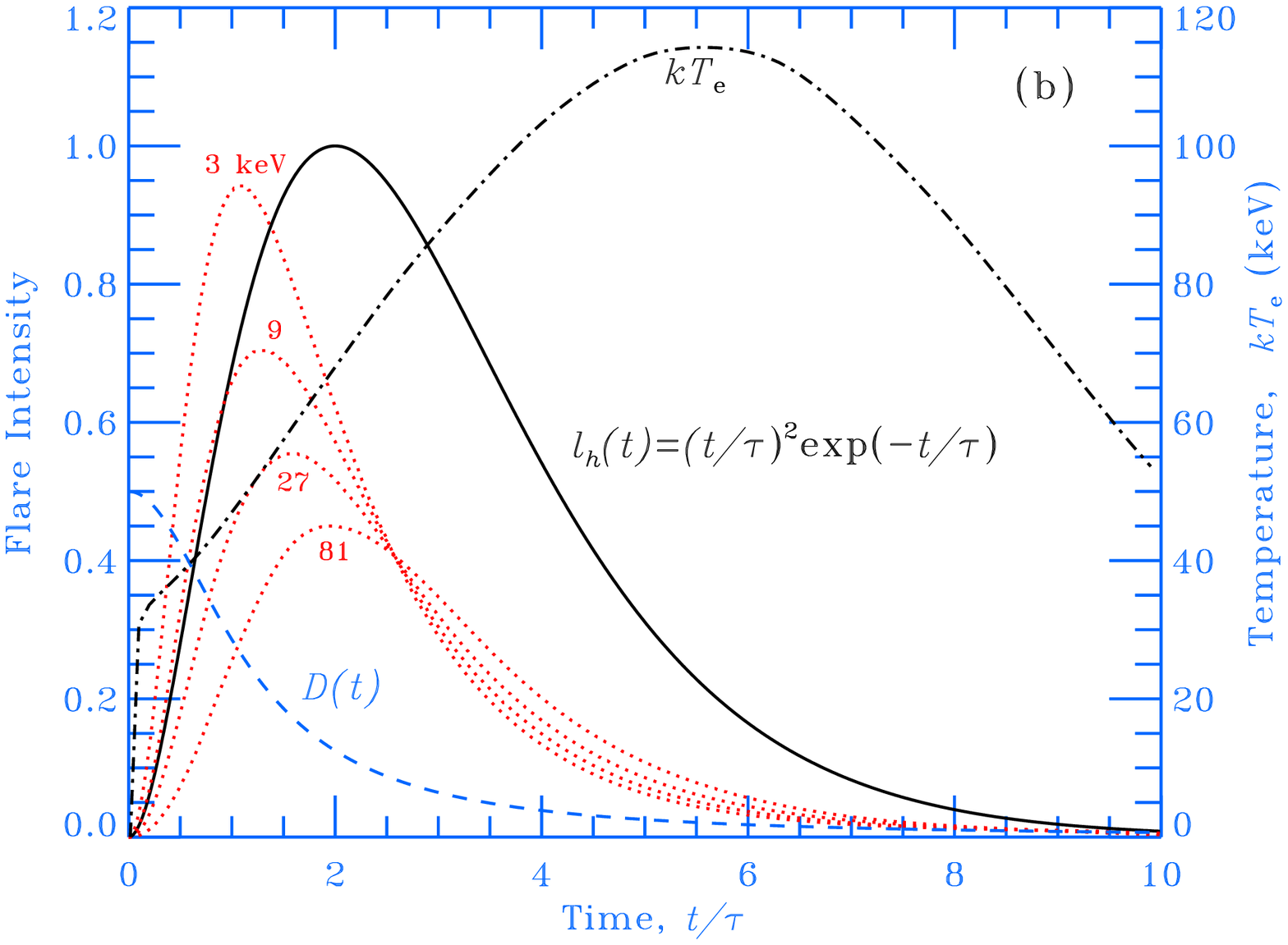,width=6.3cm,height=6cm}}
\caption{\small
(a) Spectral evolution of a magnetic flare.  Time resolved spectra (without
   Compton  reflection) are presented by {\it dotted} curves; marks are times,
   $t/\tau$,  from  the  beginning  of the  flare.  The time-averaged  Comptonized
   spectrum  is shown by a {\it thin solid}  curve.  The time-averaged  spectrum of
   Cygnus X-1 (simultaneous {\it Ginga} and OSSE data from June 1991, the data set
   \# 1 in  Gierli\'nski  et al.  1997) is plotted with crosses and the best fit
   with the  flare  model with a {\it  solid}  curve  ($\chi^2/$dof$=50.0/75$).
   Interstellar absorption is removed when plotting the model spectrum.
   (b) The flare light  curves at 3, 9, 27, and 81 keV are  presented  by {\it
   dotted}  curves.  {\it  Solid}  curve - the  heating  rate,  $\lh(t)\propto
   (t/\tau)^2\exp(-t/\tau)$;  {\it dashed} curve - the feedback factor $D(t)$;
   {\it dot-dashed} curve - the temperature of the emission region.
   See Poutanen \& Fabian (1999a,b) for details.
}
\label{fig:flare}
\end{figure}

   Miyamoto et al.  (1988) pointed out that some modulation  mechanism must
   be invoked to  produce  the  strongly  frequency-dependent  time  lags.
   Poutanen \& Fabian  (1999a,b)  proposed a model  where the time lags are
   produced by the evolution of the flare  spectrum.  They assumed that the
   energy dissipation  varies in time.  For a small (of the order of $R_g$)
   emitting  region  (ER) one can  consider  the  spectral  evolution  as a
   sequence of steady-states  as long as the  characteristic  time scale of
   variability is $\tau\gg R/c$.  Any changes in the  amplification  factor
   $A$ would cause spectral  variability  and,  specifically,  a continuous
   increase of $A$ with time during the course of the flare  would  cause a
   soft-to-hard spectral evolution producing hard time lags of the order of
   the  flare  time  scale  (see  \S~\ref{sec:shot}).  Poutanen  \&  Fabian
   (1999b) considered three mechanisms  that can increase $A$.  \\
   (1) A  flare  starts  in the  background  of  soft  photons.  The  X-ray
   spectrum is soft as long as $L_h<L_{s, \rm bkg}$.  With increasing  $L_h$, the
   soft photon input gets dominated by  reprocessed  photons.  The spectral
   slope is then determined by the feedback  parameter  $D\approx 1/A$ (the
   fraction of $L_h$ returned to the ER after the  reprocessing in the disc
   into  soft  seed  photons,  see  Stern  et  al.  1995;  Svensson   1996;
   Beloborodov 1999b) and the spectrum becomes hard.  \\
   (2) The  dissipation is accompanied by pumping net momentum into the hot
   plasma of the  emission  region  (Beloborodov  1999a,b). The resulting 
   bulk velocity increases with increasing luminosity. It leads to a lower 
   feedback and higher $A$ (if the velocity is directed away from the disc).\\
   (3) The  differential  rotation of the  footpoints of a magnetic loop at
   the disc surface  causes a twisting and elevation of the loop  (Romanova
   et al.  1998).  The time scale of the  evolution  is of the order of the
   Keplerian  time-scale.  When the  emission  region  moves  away from the
   disc,   the   feedback    decreases   and   $A$   increases    (see
   Fig.~\ref{fig:flare}).

   In all these cases the spectral  evolution  proceeds  from soft to hard.
   The {\it hard} time lags  between  energies  $E$ and $E_0$ are  $\propto
   \tau \ln  (E/E_0)$.  If there is a  distribution  of time scales  $\tau$
   between, say, 1~ms and 0.3~s (e.g.,  Keplerian  time scales at the radii
   between the innermost  radius of the accretion  disc and $\sim 50 R_g$),
   the time lags $\propto  1/f$ at the  characteristic  frequencies  of the
   variability (see \S~\ref{sec:shot} and Fig.~\ref{fig:cygx1lags}).

   If $\tau\sim$ a few light  crossing  time of the ER, the spectrum in the
   beginning of the flare is hard because of photon starvation (one needs a
   few  $R/c$ to get  reprocessed  soft  photons  into the ER) and  softens
   towards  the end of the flare  (Poutanen  \&  Fabian  1999a;  Malzac  \&
   Jourdain 2000)  producing {\it soft} time lags.  Observing the change of
   sign of the time lags at some Fourier  frequency,  $f_{\rm  sgn}$, would
   determine  the  size  of  the  ER  that  produces variability  at  these
   frequencies  (e.g.,  for Cyg  X-1, $R  \lesssim  10 R_g  (30$~Hz$/f_{\rm
   sgn})$).
   
   \vspace{0.7cm}

\subsection{Small Scale Spectral Transitions}

   The models  considered above can explain the time lags in the broad-band
   noise.  In some  QPO  sources  the  time  lags  show a very  complicated
   behaviour (see \S~\ref{sec:lags}) and may require different explanation.
   In  radiation-hydrodynamic  model (see, e.g., Lamb 1989;  Miller \& Lamb
   1992), QPOs appear as a result of  oscillations  in the optical depth of
   the radial flow due to the  radiation  feedback  from the  neutron  star
   surface.  The resulting  spectral  pivoting  produces phase lags and the
   increase of the rms  amplitude  variability  above the  pivoting  point.
   This model, however, is not applicable to the QPOs and the lags in black
   holes sources (see, e.g.,  Takizawa et al.  1997) because of the absence
   of a hard surface.

   The galactic microquasar GRS~1915+105 shows large amplitude oscillations
   with periods varying from $<$1 up to 100 s.  These time scales are a few
   orders of magnitudes  larger than Keplerian time scales and up to $10^6$
   time larger than the light crossing time of the ER.  The best  candidate
   for  producing  the  spectral   variability  that  causes  the  lags  in
   GRS~1915+105 is the  oscillation of the inner radius of the cold disc on
   viscous  time  scales.  Such  oscillations  are similar to the  spectral
   transitions,  but have  smaller  amplitude  and  occur at  shorter  time
   scales.  Changes of the relative geometry of the hot corona and the cold
   disc (with or without  changes of the total  luminosity)  cause spectral
   pivoting at a few keV (see, e.g.,  Poutanen,  Krolik, \& Ryde 1997; Esin
   et al.  1998).  The  fluxes  below and above the pivot  point  oscillate
   then  with a phase  shift of  $\sim\pi$.  The rms  amplitude  of the QPO
   increases  with the energy.  The phase lags between the  energies  above
   the  pivot  point  can then be  produced  if the  oscillations  are time
   asymmetric (see Fig.~11 in Morgan,  Remillard, \& Greiner 1997; Vilhu \&
   Nevalainen 1998).

   Similar  (but  aperiodic)  changes  in  the  inner  disc  radius  can be
   responsible  for the  broad-band  variability  observed at $f<1$~Hz  in,
   e.g., Cyg X-1.  Associated  spectral changes can manifest  themselves in
   time lags observed at these frequencies.

\subsection{Delays due to Compton Reflection and Reprocessing}

   The spectra of  accreting  GBHs and  neutron  stars show  signatures  of
   Compton reflection (see \S~\ref{sec:intro}).  Some fraction of the X-ray
   photons can be reflected from the outer edge of a flared accretion disc,
   a wind from the  companion,  etc.  Such a  reflector  acts as a low pass
   filter  smearing out the high  frequency  variations  and produces  lags
   corresponding  to the light travel time to the  reflector  only at lower
   frequencies.  Such  processes  can  explain  the  break in the  time lag
   spectra observed in Cyg~X-1 (see  Fig.~\ref{fig:cygx1lags}) and in other
   GBHs at $f\lesssim 1$ Hz.

   Reprocessed  soft  radiation  which  accompanies  Compton  reflection is
   emitted in the optical and UV spectral bands if the reprocessing  occurs
   far away from the  central  X-ray  source.  The time  delays can then be
   measured between the optical/UV and the X-ray radiation  (e.g., Hynes et
   al.  1998).  On the other  hand,  reprocessing  in the  vicinity  of the
   X-ray  emitting  region,  produces  time delays of the order of the time
   scale of the spectral evolution in the hard X-ray band.  This may be one
   reason for the  observed  soft lags in the soft state of GX 339-4 and GS
   1124-68 (\S~\ref{sec:lagbh}).

\subsection{Hot Spots on the Neutron Star Surface}

   Some neutron star sources show lags in their  periodic  oscillations  at
   kHz  frequencies.  Ford et al.  (1999) and Ford (2000)  interpreted  the
   soft lags in  Aquila~X-1  and in the  accreting  millisecond  pulsar SAX
   1808.4-3658  using a model of a  rotating  hot  spot  with a black  body
   spectrum at the surface of a neutron  star where the lags  appear due to
   Doppler effects.  The weak energy dependence of the rms amplitude of the
   oscillations  reported  by Cui et al.  (1998)  rules out the black  body
   model  for  the  spectrum  used  by  Ford.  Detailed   analysis  of  the
   pulsations in the time domain by folding techniques by Revnivtsev (1999)
   revealed  that the pulse  profile is distorted  at  different  energies,
   while the minima are reached at the same time (i.e.,  there are {\it no}
   lags in the normal meaning of this word).

\section{Summary}

   Time lags and other temporal variability data provide strong constraints
   on the models of the X-ray production.  It was demonstrated  that static
   Compton cloud models are based on  physically  unrealistic  assumptions.
   The models  invoking  spectral  evolution of the flare  spectrum can fit
   both the CCF and the time lag  Fourier  spectra  only if (1) the  energy
   dissipation  rate  increases  slowly and  decreases  rapidly and (2) the
   flare  spectrum  evolves  from soft to hard.  If soft seed  photons  are
   produced by  reprocessing  the hard ones, the change of sign in the time
   lag spectrum is expected at high frequencies  corresponding to the light
   crossing  time of the  emission  region.  The  absence  of such a change
   would put constraints on the size of the emitting region.

    We also argued that the  reflection  of hard X-rays from the outer part
   of the accretion  disc produces  time delays that we already  might have
   observed in GBHs.  If so, the disc should be flared and the break in the
   time lag Fourier  spectra then  corresponds to the size of the accretion
   disc.  Of course, such an interpretation  is not unique.  Alternatively,
   small scale spectral transitions (e.g., oscillations of the inner radius
   of the accretion  disc at viscous time scales)  might  produce time lags
   observed at lower frequencies.

   In the case of (quasi-)  periodic  oscillations  from the  neutron  star
   sources, we argued that in order to reproduce both the time lags and the
   energy dependent rms amplitude, the spectrum of the hot spots should not
   be close to a black-body.

\begin{acknowledgements}
   This work was supported by the Swedish Natural Science Research  Council
   and the Anna-Greta  and Holger  Crafoord  Fund.  I thank Katja
   Pottschmidt   for  providing  the time lag Fourier spectra and
   the  light  curves  of  Cyg  X-1  used  in
   the calculations of the cross-correlation functions.
   I am grateful to Andrei Beloborodov and Roland Svensson for valuable
   comments.
\end{acknowledgements}

\end{document}